\documentclass[5p,final,times,sort&compress]{elsarticle}
\pdfoutput=1
\usepackage{amsmath}
\usepackage{amssymb}
\usepackage{graphicx}
\usepackage[breaklinks]{hyperref}
\usepackage{textcomp}
\usepackage{url}
\usepackage[T1]{fontenc}
\usepackage[utf8]{inputenc}
\usepackage{newunicodechar}
\usepackage{fixltx2e} % \textsubscript
\usepackage{longtable}
\usepackage{mdwtab}

\newunicodechar{⁰}{\ensuremath{^0}}
\newunicodechar{¹}{\ensuremath{^1}}
\newunicodechar{²}{\ensuremath{^2}}
\newunicodechar{³}{\ensuremath{^3}}
\newunicodechar{⁴}{\ensuremath{^4}}
\newunicodechar{⁵}{\ensuremath{^5}}
\newunicodechar{⁶}{\ensuremath{^6}}
\newunicodechar{⁷}{\ensuremath{^7}}
\newunicodechar{⁸}{\ensuremath{^8}}
\newunicodechar{⁹}{\ensuremath{^9}}
\newunicodechar{⁻}{\ensuremath{^-}}
\newunicodechar{π}{\ensuremath{\pi}}
\newunicodechar{Ω}{\ensuremath{\Omega}}

\hypersetup{colorlinks=true, linkcolor=blue!50!black, urlcolor=blue!50!black, citecolor=blue!50!black}

\renewcommand{\vec}[1]{\boldsymbol{\mathbf{#1}}}

\begin{document}

\author[cenoli,cpt]{Pierre de Buyl\corref{cor}\fnref{kul}\fnref{phone}}
\ead{pdebuyl@ulb.ac.be}
\address[cenoli]{Center for Nonlinear Phenomena and Complex Systems, Université
  libre de Bruxelles, CP231, Campus Plaine, 1050 Brussels, Belgium}

\author[cpt]{Peter H. Colberg}
\address[cpt]{Chemical Physics Theory Group, Department of Chemistry,
  University of Toronto, Toronto, Ontario M5S~3H6, Canada}

\author[mpi,itp4]{Felix H{\"o}f{}ling}
\address[mpi]{Max-Planck-Institut für Intelligente Systeme, Heisenbergstraße~3,
  70569 Stuttgart, Germany}
\address[itp4]{IV. Institut für Theoretische Physik,
  Universität Stuttgart, Pfaffenwaldring~57, 70569 Stuttgart, Germany}

\cortext[cor]{Corresponding author}
\fntext[kul]{Present address: Department of Chemistry, Katholieke Universiteit Leuven, Celestijnenlaan 200F, B-3001 Heverlee, Belgium\\ Phone: +32.16.3.27355}

\title{{H5MD}: a structured, efficient, and portable file format for molecular data}

\begin{abstract}
  We propose a new file format named ``H5MD'' for storing molecular
  simulation data, such as trajectories of particle positions and velocities,
  along with thermodynamic observables that are monitored during the course of
  the simulation.
  H5MD files are HDF5 (Hierarchical Data Format) files with a specific hierarchy
  and naming scheme.
  Thus, H5MD inherits many benefits of HDF5, e.g., structured layout of
  multi-dimensional datasets, data compression, fast and parallel I/O, and
  portability across many programming languages and hardware platforms.
  H5MD files are self-contained and foster the reproducibility of scientific
  data and the interchange of data between researchers using different
  simulation programs and analysis software.
  In addition, the H5MD specification can serve for other kinds of data (e.g.
  experimental data) and is extensible to supplemental data, or may be part of
  an enclosing file structure.
\end{abstract}
\maketitle

\section{Introduction}

Storing molecular data, such as particle coordinates, is a problem faced by
simulation scientists in statistical physics, physical chemistry, structural
biology, crystallography, among others.
Although there are currently various simulation packages that are related to
specific fields of research and that define their own file formats, there is no
standard file format for the storage and communication of molecular data.

The popular PDB file format of the Protein Data Bank~\cite{pdb} formats is
accepted as input to several molecular dynamics programs
(Gromacs~\cite{van_der_spoel_et_al_gromas_jcc_2005},
NAMD~\cite{phillips_et_al_namd_jcc_2005}, CHARMM~\cite{CHARMM2009});
the original purpose of PDB, however, was to store reference crystallographic
structures from experiments.
PDB is not very practical as it is text-file based and uses a rigid
organization.
While a text file is simple to write, it uses more disk space than a binary
file, it imposes a sequential reading of the data which can be time and
memory intensive, and it prevents modularity of the stored data.
Such technical considerations did not prevent researchers to base their work
and software on PDB for running high-performance simulations and performing
data analysis, in addition to using PDB for crystallographic data.
For example, PDB focuses on the storage of simulation data at a single instance
in time, but non-standard extensions of PDB are sometimes used and allow one to
store several time frames.
The documentations of the MDAnalysis software~\citep[Sec.~5.8]{mdanalysis_2011}
and of trjconv, a utility of Gromacs~\cite{van_der_spoel_et_al_gromas_jcc_2005},
mention explicitly this possibility.
Other molecular dynamics packages use differently structured text files
(ESPResSo~\cite{arnold13a, limbach06a}, LAMMPS~\cite{LAMMPS2013}) or resort to
XML as a structured, but still text-based format (HOOMD \cite{HOOMD2013}).

Many technical features are nowadays available for file formats that are highly
desirable for storing scientific data; examples include self-descriptiveness,
native support of data arrays, portability across hardware platforms, and
availability of computer libraries for many programming languages.
The absence of a reference file format of satisfactory technical quality does a
disservice to the various communities that generate and analyze molecular data.
Insights on this problem are presented in a recent publication by
Hinsen~\cite{hinsen_cise_2012}.
Discussions on file formats for molecular data have also been held on the
mailing list of Gromacs~\cite{GROMACS-mail}, suggesting that the need
for technical solutions is recognized among experts in the field of molecular
simulation.

Designing a new file format provides the opportunity to take into account
progress in software engineering.
Metadata on the person that authored a file, on the software that wrote it,
and on the creation date can be included, for instance.
This supports an approach of reproducible research with clear data provenance
and definition~\cite{peng_reproducible_research_2011}.
On the way towards reproducible simulation-based research, the availability of
source code is only one part of an ensemble of
requirements~\cite{ince_et_al_open_programs_nature_2012}.
Another necessary building block in this process is the design of
well-documented, published, royalty-free file formats.

Here we propose a file format named H5MD to store molecular data.
It is based on the Hierarchical Data Format version 5 (HDF5)~\cite{HDF5} as an
underlying technology.
H5MD specifies how to store and organize data for the trajectory of a complex
many-particle system and for time-dependent or -independent observables.
In addition, H5MD metadata are defined, and there is scope to store custom data
such as flow fields of an embedding solvent, control parameters, or simulation scripts.
The H5MD specification is available under the terms of the GNU General Public
License version 3 (or any later version)~\cite{GPLv3}, the development is
hosted on the open-source platform Savannah~\cite{savannah}, and discussions
have been held from the early stages of the project on a public mailing
list~\cite{h5md-user}.
With the objectives to publish the specification, to ease the use of H5MD, to
share good practice and practical information on the development of software
using H5MD, we also use the website feature of the Savannah platform, see
Ref.~\cite{h5md-website}.
H5MD could find applications in simulation programs (molecular dynamics, fluid
dynamics, etc.), experimental data acquisition (e.g., particle tracking,
particle imaging velocimetry), and the analysis tools related to these data.

An outline of the article follows.
Section~\ref{design} discusses the requirements faced with regard to
storage of scientific data, leading to the choice of HDF5 as the basis of H5MD.
Section~\ref{h5md} presents the H5MD format specification version 1.0.0.
Section~\ref{use} presents use cases from simulation software that implement
H5MD and a Python library that is accompanied by illustrative examples.

\section{Design considerations}
\label{design}

H5MD is developed with several design goals in mind:

\begin{description}
\item[Versatility] H5MD does not adhere to a specific application or research
  project.
  Rather it adopts an abstract language to define generic data elements and
  structures common to molecular data.
  H5MD files can be used to store data from particle-based simulations, but
  also molecular structures from experimental measurements, or they can serve
  as input for subsequent investigations.

\item[Modularity] H5MD allows one to store heterogeneous data independently of
  each other within one file to reflect several facets of a complex simulation.
  For example, the particle trajectory may be combined with the sampling of
  thermodynamic observables during the simulation.
  These data may be split further into subsets of a complex system, e.g., the
  atoms forming colloidal nanoparticles and the surrounding solvent molecules.

\item[Extensibility] H5MD files may store additional data that is not part of
  the specification without breaking its H5MD compliance.
  Storing application-specific data such as control parameters or simulation
  scripts fosters the reproducibility of scientific results.

\item[Self-descriptiveness] A data element stored within H5MD carries a
  descriptive name and specifies its data type and array shape.
  Creation and modification of data elements can be tracked.
  Data elements can be annotated further by custom metadata.

\item[Efficiency] H5MD files allow for high-throughput I/O and compression to
  reduce storage requirements.
  Further, random access to data subsets is fast and (nearly) constant in time.

\item[Portability] H5MD files can be read and written across many different
  operating systems and computer architectures.
  Access to H5MD files can be achieved easily by a large variety of programming
  languages.
  In particular, H5MD can be integrated natively into existing software.
\end{description}

The goals ``Modularity'', ``Extensibility'', and
``Self{-\penalty0\hskip0pt}descriptive-ness'' call for a structured file format.
A binary and structured file format is demanded by the requirement
``Efficiency'', and ``Portability'' is ensured by resorting to an I/O library
that is well supported by and established in the scientific community.

The HDF5 file format~\cite{HDF5} is an ideal technology to meet these
requirements.
HDF5 is a mature, structured, binary, and portable file format for storing
scientific data in a self-describing manner and consists of a specification and
an I/O library.
Data is stored in named datasets that are organized in groups similarly to the
tree structure of a filesystem.
Another significant benefit over non-binary or non-structured file formats are
the facilities of HDF5 for parallel input/output, for data compression, and for
random access to parts of a multi-dimensional dataset without reading the full
dataset (array slicing).
The HDF5 library provides programming interfaces for C, C++, and Fortran; Java
and Python bindings are available, and some commercial packages (e.g., Matlab,
Mathematica, IDL) support HDF5.
Still there remains a great deal of choices to be made with respect to the
organization of the file, the naming scheme, the coherence of time-dependent
information, etc.
Those choices form the H5MD specification.

The authors took care to respect HDF5 technicalities~\cite{HDF5_advanced_topics,
HDF5_reference_manual}.
Creation and modification times in a H5MD file are recorded implicitly using
the object tracking feature of HDF5.
Compact Datasets are used for small amounts of data (i.e., when the data size
is independent of the system size).
Data forming a time series (e.g., particle positions in the course of a
simulation) is stored as a single dataset, where the time axis is accommodated
by an extra, extensible dimension prepended to the array shape.
This approach is preferred over storing the sequence of frames in a sequence of
differently named datasets (or groups), avoiding \emph{inter alia} the
clobbering of the file structure for a large number of frames.
Since molecular data are sometimes time-dependent and sometimes
time-independent, H5MD users may choose for any data item among two
well-defined, generic data structures.

The design of H5MD also strives to be future-proof in that it allows for future
extensions and avoids setting in stone features that appear not yet mature or
generic enough.
For instance, the storage of protein topology or crystallographic symmetry
groups are not part of the specification though the authors recognize their
importance.
The self-describing and organized nature of HDF5 allows additions in the form
of new (root) groups or datasets without breaking the compliance of the file
with H5MD.
On the other hand, the H5MD format is defined as an HDF5 group within
an HDF5 file and a single HDF5 file may thus contain several H5MD root groups,
allowing for the collection of data in a single file and for making H5MD
part of an enclosing file structure, e.g., in the Mosaic data model~\cite{mosaic,hinsen_mosaic_2014}.

Some features or constraints are included in H5MD by so-called \emph{modules}.
Modules are aimed at specific applications or domains in which agreed
upon conventions exist and should be reflected in the data.
Two such modules are presented in appendix~\ref{modules}.
The module ``Units'', for instance, describes how physical units can be
meaningfully attached to H5MD data.

\section{H5MD format specification version 1.0.0}
\label{h5md}

\subsection{Objective}

H5MD stands for ``HDF5 for molecular data''. H5MD is a specification to
store molecular simulation data and is based on the
\href{http://www.hdfgroup.org/HDF5/}{HDF5} file format \citep{HDF5}. The
primary goal is to facilitate the portability of said data amongst
scientific simulation and analysis programs.

\subsection{File format}

H5MD structures are stored in the
\href{http://www.hdfgroup.org/HDF5/doc/H5.format.html}{HDF5 file format}
version 0 or later. It is recommended to use the HDF5 file format
version 2, which includes the implicit tracking of the creation and
modification times of the file and of each of its objects.

\subsection{Notation and naming}

HDF5 files are organized into groups and datasets, summarized as
\emph{objects}, which form a tree structure with the datasets as leaves.
Attributes can be attached to each object. The H5MD specification adopts
this naming and uses the following notation to depict the tree or its
subtrees:

\begin{description}
\item[\texttt{\textbackslash{}-{}- item}]
An object within a group, that is either a dataset or a group. If it is
a group itself, the objects within the group are indented by five spaces
with respect to the group name.
\item[\texttt{+-{}- attribute}]
An attribute, that relates either to a group or a dataset.
\item[\texttt{\textbackslash{}-{}- data: \textless{}type\textgreater{}{[}dim1{]}{[}dim2{]}}]
A dataset with array dimensions \texttt{dim1} by \texttt{dim2} and of
type \texttt{\textless{}type\textgreater{}}. The type is taken from
\texttt{Enumeration}, \texttt{Integer}, \texttt{Float} or
\texttt{String} and follows the HDF5 Datatype classes. If the type is
not mandated by H5MD, \texttt{\textless{}type\textgreater{}} is
indicated. A scalar dataspace is indicated by \texttt{{[}{]}}.
\item[\texttt{(identifier)}]
An optional item.
\item[\texttt{\textless{}identifier\textgreater{}}]
An optional item with unspecified name.
\end{description}

H5MD defines a structure called \emph{H5MD element} (or \emph{element}
whenever there is no confusion). An element is either a time-dependent
group or a single dataset (see time-dependent data below), depending on
the situation.

\subsection{General organization}

H5MD defines an organization of the HDF5 file or a part thereof into
groups, datasets, and attributes. The root level of the H5MD structure
may coincide with the root of the HDF5 file or be an arbitrary group
inside the HDF5 tree. A number of groups are defined at the H5MD root
level. Several levels of subgroups may exist inside the H5MD structure,
allowing the storage and description of subsystems.

The H5MD structure is allowed to possess non-specified groups, datasets,
or attributes that contain additional information such as
application-specific parameters or data structures, leaving scope for
future extensions. Only the \texttt{h5md} group is mandatory at the H5MD
root level. All other root groups are optional, allowing the user to
store only relevant data. Inside each group, every group or dataset is
again optional, unless specified differently.

H5MD supports equally the storage of time-dependent and time-independent
data, i.e., data that change in the course of the simulation or that do
not. The choice between those storage types is not made explicit for the
elements in the specification, it has to be made according to the
situation. For instance, the species and mass of the particles are often
fixed in time, but in chemically reactive systems this might not be
appropriate.

\subsubsection{Time-dependent data}

Time-dependent data consist of a series of samples (or frames) referring
to multiple time steps. Such data are found inside a single dataset and
are accessed via dataset slicing. In order to link the samples to the
time axis of the simulation, H5MD defines a \emph{time-dependent H5MD
element} as a group that contains, in addition to the actual data,
information on the corresponding integer time step and on the physical
time. The structure of such a group is:

\begin{verbatim}
<element>
 \-- step: Integer[variable]
 \-- time: Float[variable]
 \-- value: <type>[variable][...]
\end{verbatim}

\begin{description}
\item[\texttt{step}]
A dataset with dimensions \texttt{{[}variable{]}} that contains the time
steps at which the corresponding data were sampled. It is of
\texttt{Integer} type to allow exact temporal matching of data from one
H5MD element to another. The values of the dataset are in monotonically
increasing order.
\item[\texttt{time}]
A dataset that is the same as the \texttt{step} dataset, except it is
\texttt{Float}-valued and contains the simulation time in physical
units. The values of the dataset are in monotonically increasing order.
\item[\texttt{value}]
A dataset that holds the data of the time series. It uses a simple
dataspace whose rank is given by 1 plus the tensor rank of the data
stored. Its shape is the shape of a single data item prepended by a
\texttt{{[}variable{]}} dimension that allows the accumulation of
samples during the course of time. For instance, the data shape of
scalars has the form \texttt{{[}variable{]}}, \texttt{D}-dimensional
vectors use \texttt{{[}variable{]}{[}D{]}}, etc. The first dimension of
\texttt{value} must match the unique dimension of \texttt{step} and
\texttt{time}.
\end{description}

If several H5MD elements are sampled at equal times, \texttt{step} and
\texttt{time} of one element may be hard links to the \texttt{step} and
\texttt{time} datasets of a different element. If two elements are
sampled at different times (for instance, if one needs the positions
more frequently than the velocities), \texttt{step} and \texttt{time}
are unique to each of them.

\subsubsection{Time-independent data}

H5MD defines a \emph{time-independent H5MD element} as a dataset. As for
the \texttt{value} dataset in the case of time-dependent data, data type
and array shape are implied by the stored data, where the
\texttt{{[}variable{]}} dimension is omitted.

\subsubsection{Storage order of arrays}

All arrays are stored in C-order as enforced by the HDF5 file format
(see
\href{http://www.hdfgroup.org/HDF5/doc/UG/12_Dataspaces.html\#ProgModel}{§
3.2.5} in \citep{HDF5_users_guide}). A C or C++ program may thus declare
\texttt{r{[}N{]}{[}D{]}} for the array of particle coordinates while the
Fortran program will declare a \texttt{r(D,N)} array (appropriate index
ordering for a system of \texttt{N} particles in \texttt{D} spatial
dimensions), and the HDF5 file will be the same.

\subsection{H5MD root level}

The root level of an H5MD structure holds a number of groups and is
organized as follows:

\begin{verbatim}
H5MD root
 \-- h5md
 \-- (particles)
 \-- (observables)
 \-- (parameters)
\end{verbatim}

\begin{description}
\item[\texttt{h5md}]
A group that contains metadata and information on the H5MD structure
itself. It is the only mandatory group at the root level of H5MD.
\item[\texttt{particles}]
An optional group that contains information on each particle in the
system, e.g., a snapshot of the positions or the full trajectory in
phase space. The size of the stored data scales linearly with the number
of particles under consideration.
\item[\texttt{observables}]
An optional group that contains other quantities of interest, e.g.,
physical observables that are derived from the system state at given
points in time. The size of stored data is typically independent of the
system size.
\item[\texttt{parameters}]
An optional group that contains application-specific, custom data such
as control parameters or simulation scripts.
\end{description}

In subsequent sections, the examples of HDF5 organization may start at
the group level, omitting the display of \texttt{H5MD root}.

\hyperdef{}{h5md-metadata}{\subsection{H5MD
metadata}\label{h5md-metadata}}

A set of global metadata describing the H5MD structure is stored in the
\texttt{h5md} group as attributes. The contents of the group is:

\begin{verbatim}
h5md
 +-- version: Integer[2]
 \-- author
 |    +-- name: String[]
 |    +-- (email: String[])
 \-- creator
      +-- name: String[]
      +-- version: String[]
\end{verbatim}

\begin{description}
\item[\texttt{version}]
An attribute, of \texttt{Integer} datatype and of simple dataspace of
rank 1 and size 2, that contains the major version number and the minor
version number of the H5MD specification the H5MD structure conforms to.

The version \emph{x.y.z} of the H5MD specification follows
\href{http://semver.org/spec/v2.0.0.html}{semantic versioning}
\citep{semantic_versioning}: A change of the major version number
\emph{x} indicates backwards-incompatible changes to the file structure.
A change of the minor version number \emph{y} indicates
backwards-compatible changes to the file structure. A change of the
patch version number \emph{z} indicates changes that have no effect on
the file structure and serves to allow for clarifications or minor text
editing of the specification.

As the \emph{z} component has no impact on the content of a H5MD file,
the \texttt{version} attribute contains only \emph{x} and \emph{y}.
\item[\texttt{author}]
A group that contains metadata on the person responsible for the
simulation (or the experiment) as follows:

\begin{description}
\item[\texttt{name}]
An attribute, of fixed-length string datatype and of scalar dataspace,
that holds the author's real name.
\item[\texttt{email}]
An optional attribute, of fixed-length string datatype and of scalar
dataspace, that holds the author's email address of the form
\texttt{email@domain.tld}.
\end{description}
\item[\texttt{creator}]
A group that contains metadata on the program that created the H5MD
structure as follows:

\begin{description}
\item[\texttt{name}]
An attribute, of fixed-length string datatype and of scalar dataspace,
that stores the name of the program.
\item[\texttt{version}]
An attribute, of fixed-length string datatype and of scalar dataspace,
that yields the version of the program.
\end{description}
\end{description}

\subsubsection{Modules}

The H5MD specification can be complemented by modules specific to a
domain of research. A module may define additional data elements within
the H5MD structure, add conditions that the data must satisfy, or define
rules for their semantic interpretation. Multiple modules may be
present, as long as their prescriptions are not contradictory. Each
module is identified by a name and a version number.

The modules that apply to a specific H5MD structure are stored as
subgroups within the group \texttt{h5md/modules}. Each module holds its
version number as an attribute, further module-specific information may
be stored:

\begin{verbatim}
h5md
 \-- (modules)
      \-- <module1>
      |    +-- version: Integer[2]
      \-- <module2>
      |    +-- version: Integer[2]
      \-- ...
\end{verbatim}

\begin{description}
\item[\texttt{version}]
An attribute, of \texttt{Integer} datatype and of simple dataspace of
rank 1 and size 2, that contains the major version number and the minor
version number of the module.

The version \emph{x.y.z} of an H5MD module follows
\href{http://semver.org/spec/v2.0.0.html}{semantic versioning}
\citep{semantic_versioning} and again only the components \emph{x} and
\emph{y} are stored, see \texttt{h5md/version} in
``\hyperref[h5md-metadata]{H5MD metadata}.''
\end{description}

\subsection{Particles group}

Information on each particle, i.e., particle trajectories, is stored in
the \texttt{particles} group. The \texttt{particles} group is a
container for subgroups that represent different subsets of the system
under consideration, and it may hold one or several subgroups, as
needed. These subsets may overlap and their union may be incomplete,
i.e., not represent all particles of the simulation volume. The
subgroups contain the trajectory data for each particle as
time-dependent or time-independent data, depending on the situation.
Each subgroup contains a specification of the simulation box, see below.
For each dataset, the particle index is accommodated by the second
(first, in the case of time-independence) array dimension.

The contents of the \texttt{particles} group assuming \texttt{N}
particles in \texttt{D}-dimensional space could be the following:

\begin{verbatim}
particles
 \-- <group1>
      \-- box
      \-- (position)
      |    \-- step: Integer[variable]
      |    \-- time: Float[variable]
      |    \-- value: <type>[variable][N][D]
      \-- (image)
      |    \-- step: Integer[variable]
      |    \-- time: Float[variable]
      |    \-- value: <type>[variable][N][D]
      \-- (species: Enumeration[N])
      \-- ...
\end{verbatim}

The following identifiers for H5MD elements are standardized:

\begin{description}
\item[\texttt{position}]
An element that describes the particle positions as coordinate vectors
of \texttt{Float} or \texttt{Integer} type.

If the component $k$ of \texttt{box/boundary} (see
\hyperref[simulation-box]{below}) is set to \texttt{none}, the data
indicate for each particle the component $k$ of its absolute position in
space. If the component $k$ of \texttt{box/boundary} is set to
\texttt{periodic}, the data indicate for each particle the component $k$
of the absolute position in space of an \emph{arbitrary} periodic image
of that particle.
\item[\texttt{image}]
An element that represents periodic images of the box as coordinate
vectors of \texttt{Float} or \texttt{Integer} type and allows one to
compute for each particle its absolute position in space. If
\texttt{image} is present, \texttt{position} must be present as well.
For time-dependent data, the \texttt{step} and \texttt{time} datasets of
\texttt{image} must equal those of \texttt{position}, which must be
accomplished by hard-linking the respective datasets.

If the component $k$ of \texttt{box/boundary} (see
\hyperref[simulation-box]{below}) is set to \texttt{none}, the values of
the corresponding component $k$ of \texttt{image} serve as placeholders.
If the component $k$ of \texttt{box/boundary} is set to
\texttt{periodic}, for a cuboid box, the component $k$ of the absolute
position of particle $i$ is computed as $R_{ik} = r_{ik} + L_k a_{ik}$,
where $\vec r_i$ is taken from \texttt{position}, $\vec a_i$ is taken
from \texttt{image}, and $\vec L$ from \texttt{box/edges}.
\item[\texttt{velocity}]
An element that contains the velocities for each particle as a vector of
\texttt{Float} or \texttt{Integer} type.
\item[\texttt{force}]
An element that contains the total forces (i.e., the accelerations
multiplied by the particle mass) for each particle as a vector of
\texttt{Float} or \texttt{Integer} type.
\item[\texttt{mass}]
An element that holds the mass for each particle as a scalar of
\texttt{Float} type.
\item[\texttt{species}]
An element that describes the species for each particle, i.e., its
atomic or chemical identity, as a scalar of \texttt{Enumeration} or
\texttt{Integer} data type. Particles of the same species are assumed to
be identical with respect to their properties and unbonded interactions.
\item[\texttt{id}]
An element that holds a scalar identifier for each particle of
\texttt{Integer} type, which is unique within the given particle
subgroup. The \texttt{id} serves to identify particles over the course
of the simulation in the case when the order of the particles changes,
or when new particles are inserted and removed. If \texttt{id} is
absent, the identity of the particles is given by their index in the
\texttt{value} datasets of the elements within the same subgroup.

A \emph{fill value} (see
\href{http://www.hdfgroup.org/HDF5/doc/UG/11_Datatypes.html\#Fvalues}{§
6.6} in \citep{HDF5_users_guide}) may be defined for \texttt{id/value}
upon dataset creation. When the identifier of a particle is equal to
this user-defined value, the particle is considered non-existing, the
entry serves as a placeholder. This permits the storage of subsystems
whose number of particles varies in time. For the case of varying
particle number, the dimension denoted by \texttt{{[}N{]}} above may be
variable.
\end{description}

\hyperdef{}{simulation-box}{\subsection{Simulation
box}\label{simulation-box}}

The specification of the simulation box is stored in the group
\texttt{box}, which must be contained within each of the subgroups of
the \texttt{particles} group. Storing the box information at several
places reflects the fact that different subgroups may be sampled at
different time grids. This way, the box information remains associated
to a group of particles. A specific requirement for \texttt{box} groups
inside \texttt{particles} is that the \texttt{step} and \texttt{time}
datasets exactly match those of the corresponding \texttt{position}
groups, which must be accomplished by hard-linking the respective
datasets.

The spatial dimension and the boundary conditions of the box are stored
as attributes to the \texttt{box} group, e.g., :

\begin{verbatim}
particles
 \-- <group1>
      \-- box
           +-- dimension: Integer[]
           +-- boundary: String[D]
           \-- (edges)
\end{verbatim}

\begin{description}
\item[\texttt{dimension}]
An attribute that stores the spatial dimension \texttt{D} of the
simulation box and is of \texttt{Integer} datatype and scalar dataspace.
\item[\texttt{boundary}]
An attribute, of fixed-length string datatype and of simple dataspace of
rank 1 and size \texttt{D}, that specifies the boundary condition of the
box along each dimension. The values in \texttt{boundary} are either
\texttt{periodic} or \texttt{none}:

\texttt{periodic} The simulation box is periodically continued along the
given dimension and serves as the unit cell for an infinite tiling of
space.

\texttt{none} No boundary condition is imposed. This summarizes the
situations of open systems (i.e., an infinitely large box) and closed
systems (e.g., due to an impenetrable wall). For those components where
\texttt{boundary} is set to \texttt{none}, the corresponding values of
\texttt{edges} serve as placeholders.
\end{description}

Information on the geometry of the box edges is stored as an H5MD
element, allowing for the box to be fixed in time or not. Supported box
shapes are the cuboid and triclinic unit cell, for other shapes a
transformation to the triclinic shape may be considered
\citep{Bekker:1997}. If all values in \texttt{boundary} are
\texttt{none}, \texttt{edges} may be omitted.

\begin{description}
\item[\texttt{edges}]
A \texttt{D}-dimensional vector or a \texttt{D} × \texttt{D} matrix,
depending on the geometry of the box, of \texttt{Float} or
\texttt{Integer} type. If \texttt{edges} is a vector, it specifies the
space diagonal of a cuboid-shaped box. If \texttt{edges} is a matrix,
the box is of triclinic shape with the edge vectors given by the rows of
the matrix.

For a time-dependent box, a cuboid geometry is encoded by a dataset
\texttt{value} (within the H5MD element) of rank 2 (1 dimension for the
time and 1 for the vector) and a triclinic geometry by a dataset
\texttt{value} of rank 3 (1 dimension for the time and 2 for the
matrix).

For a time-independent box, a cuboid geometry is encoded by a dataset
\texttt{edges} of rank 1 and a triclinic geometry by a dataset of rank
2.
\end{description}

For instance, a cuboid box that changes in time would appear as:

\begin{verbatim}
particles
 \-- <group1>
      \-- box
           +-- dimension: Integer[]
           +-- boundary: String[D]
           \-- edges
                \-- step: Integer[variable]
                \-- time: Float[variable]
                \-- value: <type>[variable][D]
\end{verbatim}

where \texttt{dimension} is equal to \texttt{D}. A triclinic box that is
fixed in time would appear as:

\begin{verbatim}
particles
 \-- <group1>
      \-- box
           +-- dimension: Integer[]
           +-- boundary: String[D]
           \-- edges: <type>[D][D]
\end{verbatim}

where \texttt{dimension} is equal to \texttt{D}.

\subsection{Observables group}

Macroscopic observables, or more generally, averages of some property
over many particles, are stored in the root group \texttt{observables}.
Observables representing only a subset of the particles may be stored in
appropriate subgroups similarly to the \texttt{particles} tree. Each
observable is stored as an H5MD element. The shape of the corresponding
dataset (the element itself for time-independent data and \texttt{value}
for time-dependent data) is the tensor shape of the observable,
prepended by a \texttt{{[}variable{]}} dimension for time-dependent
data.

The contents of the observables group has the following structure:

\begin{verbatim}
observables
 \-- <observable1>
 |    \-- step: Integer[variable]
 |    \-- time: Float[variable]
 |    \-- value: <type>[variable]
 \-- <observable2>
 |    \-- step: Integer[variable]
 |    \-- time: Float[variable]
 |    \-- value: <type>[variable][D]
 \-- <group1>
 |    \-- <observable3>
 |         \-- step: Integer[variable]
 |         \-- time: Float[variable]
 |         \-- value: <type>[variable][D][D]
 \-- <observable4>: <type>[]
 \-- ...
\end{verbatim}

\subsection{Parameters group}

The \texttt{parameters} group stores application-specific, custom data
such as control parameters or simulation scripts. The group consists of
groups, datasets, and attributes; the detailed structure, however, is
left unspecified.

The contents of the \texttt{parameters} group could be the following:

\begin{verbatim}
parameters
 +-- <user_attribute1>
 \-- <user_data1>
 \-- <user_group1>
 |    \-- <user_data2>
 |    \-- ...
 \-- ...
\end{verbatim}

\section{Uses and tools for H5MD}
\label{use}

H5MD evolved from custom HDF5 file structures that were used by the authors in
their simulation-based research.
After discussing the common need for a file format and deciding to join
efforts, the design of H5MD started from the experience of ``HAL's MD package''
and the RMPCDMD software.

``HAL's MD package''~\cite{HALMD} exploits the vastly parallel architecture of
modern graphics processors (GPUs) and the CUDA framework (Nvidia Corp., Santa
Clara, CA) to deliver accurate and efficient, highly-accelerated large-scale
molecular dynamics simulations~\cite{Colberg2011}.
The software has been written in C++ and Lua by P.H.C.\ and F.H., and the
package is now maintained by F.H.\ It features a modular design along with a
scripting user interface and permits the simulation of some
10\textsuperscript{6} particles on a single GPU.
The focus on the online evaluation of thermodynamic observables and of dynamic
correlation functions minimizes the need to dump particle-resolved data to the
disk.
The software targets problems in statistical physics, applications include
the slow glassy dynamics of binary mixtures~\cite{Colberg2011} and the
structure of liquid--vapor interfaces~\cite{Hoefling2013}.

RMPCDMD is a Fortran hydrodynamics simulation code developed by P.d.B.\
based on the Multi-particle Collision Dynamics algorithm used for solvent
dynamics, and the code is able to perform molecular dynamics of embedded
colloids.
In addition, RMPCDMD allows the simulation of reactive fluids.
RMPCDMD uses H5MD to store the particle coordinates of the colloids and the
solvent (for checkpointing only, up to 2 millions solvent particles tested) and
all time-dependent data in the simulation.
The H5MD code of RMPCDMD is separate from the simulation code and is provided
as the f90h5md~\cite{f90h5md} Fortran module, by the same author.

In order to demonstrate the use of H5MD in a minimal self-contained example,
we provide a code for the simulation of a 1D random walk of an ensemble of
particles.
The example is written in Python using the packages NumPy, h5py~\cite{h5py2008},
and matplotlib, which are shipped with major GNU/Linux distributions or
scientific Python distributions.
The code is composed of the module pyh5md~\cite{pyh5md} (available as
supplementary material and at the Python Package Index,
\href{http://pypi.python.org/}{http://pypi.python.org/}) that provides H5MD
functionality on top of h5py, and the simulation script
\texttt{random\_walk\_1d.py}.
The management of H5MD data is handled by pyh5md, which results in a very small
number of instructions in the actual program: opening the file, declaring each
data element (particles group with position dataset, and observables), and
writing the data.
Excluding comments and blank lines, \texttt{random\_walk\_1d.py} contains
18~lines of code.
The code propagates particles and stores their positions in the course of time in
an H5MD file.
The center of mass of the system is stored as an observable in the file.
After running the simulation, an analysis of the mean-squared displacement is
presented in \texttt{random\_walk\_1d\_analysis.py}.
It illustrates the opening of a file and the acquisition of trajectory data in
about 30~lines of code.

\section{Conclusions and perspectives}

To summarize, we propose H5MD as a new file format to store molecular data.
We have also presented a simple implementation of H5MD to serve as an
illustration and a starting point for potential users.
This simple implementation, along with the fact that H5MD is employed by all
authors in their simulation codes, shows the applicability of the file format
to real-world use cases.

It is our hope that scientists from related fields will participate in the
further development of H5MD.
The modular nature of H5MD and its open development process allow for future
extensions that make the file format suitable for a wide range of applications.
In particular, we believe that H5MD has the potential to provide a superb
alternative to the PDB file format.
H5MD provides the basis for storing particle-based data that form an important
part of PDB (specifically, the ``ATOM'' records).
Additional data, many of which consist in named parameters, would fit easily as
named datasets or attributes in a HDF5 file.
Other interesting extensions are related to the storage of grid data, for
example when mesoscopic particles are coupled to flow fields as in
computational fluid dynamics or lattice-Boltzmann approaches.

\section*{Acknowledgments}

We are grateful to Konrad Hinsen for his numerous contributions on the H5MD
mailing list and to Olaf Lenz for correspondence, discussions, and comments on
the manuscript.
We acknowledge Savannah (the free software forge) for hosting the H5MD project.
We thank Bryan
Robertson and Cyrille Lavigne for critically reading the manuscript.
P.d.B.\ and P.H.C.\ wish to thank Raymond Kapral for his support.

\appendix
\def\appendixname{}
\onecolumn

\section{Modules}
\label{modules}

\subsection{Thermodynamic observables}

\subsubsection{Objective}

This module defines a set of thermodynamic observables commonly output
by molecular simulation programs.

\subsubsection{Module name and version}

The name of this module is \texttt{thermodynamics}. The module version
is 1.0.0.

\subsubsection{Observables}

Thermodynamic observables are stored in the \texttt{observables} group
for global properties or in
\texttt{observables/\textless{}group\textgreater{}} for subsystems,
similarly to the \texttt{particles} group. The groups have the following
contents:

\begin{verbatim}
observables
 \-- <group>
      +-- dimension: Integer[]
      \-- particle_number
      \-- (pressure)
      \-- (temperature)
      \-- (density)
      \-- ...
\end{verbatim}

\begin{description}
\item[\texttt{dimension}]
A scalar attribute of \texttt{Integer} type that gives the dimension of
the space embedding the subsystem.
\item[\texttt{particle\_number}]
The number of particles in the subsystem stored as scalar H5MD element
of \texttt{Integer} datatype.
\end{description}

The following H5MD elements are optional, of scalar character, and use
the \texttt{Float} datatype.

\begin{description}
\item[\texttt{pressure}]
The pressure of the subsystem.
\item[\texttt{temperature}]
The (instantaneous) temperature of the subsystem as inferred from the
kinetic energy.
\item[\texttt{density}]
The number density of the subsystem stored as \texttt{Float} or
\texttt{Integer} datatype.
\item[\texttt{potential\_energy}]
The potential energy of the subsystem.
\item[\texttt{kinetic\_energy}]
The kinetic energy of the subsystem.
\item[\texttt{internal\_energy}]
The internal energy of the subsystem, typically the sum of potential and
kinetic energy.
\item[\texttt{enthalpy}]
The enthalpy of the subsystem.
\end{description}

The latter 4 quantities are stored as per-particle averages. The
corresponding extensive quantities (scaling linearly with the system
size) are obtained by multiplication with the number of particles.
Per-volume averages follow by multiplication with the number density if
present.

\subsection{Units}

\subsubsection{Objective}

This module defines how physical units are attached to dimensionful H5MD
elements.

\subsubsection{Module name and version}

The name of this module is \texttt{units}. The module version is 1.0.0.

\subsubsection{Unit system definition}

The \texttt{units} group possesses, in addition to the \texttt{version}
attribute, a \texttt{system} attribute that defines the unit system in
use. \texttt{system} is of scalar dataspace and fixed-length string
datatype.

\subsubsection{Unit attribute}

The datasets of any H5MD element that have a physical dimension may
carry an attribute \texttt{unit} to indicate the physical unit of the
respective data. In general, this refers to the dataset itself for
time-independent elements, or to the datasets \texttt{value} and
\texttt{time} in the time-dependent case:

\begin{verbatim}
<element>
 \-- step: Integer[variable]
 \-- time: Float[variable]
 |    +-- (unit: String[])
 \-- value: <type>[variable][...]
      +-- (unit: String[])
\end{verbatim}

The attribute \texttt{unit} is of scalar dataspace and fixed-length
\texttt{String} datatype using the ASCII character set.

\subsubsection{Unit string}

The \texttt{unit} string consists of a sequence of unit factors
separated by a space. A unit factor is either a number (an integer or a
decimal fraction) or a unit symbol optionally followed by a non-zero,
signed integer indicating the power to which this factor is raised. Each
unit symbol may occur only once. There may also be at most one numeric
factor, which must be the first one.

Examples:

\begin{itemize}
\item
  ``nm+3'' stands for cubic nanometers
\item
  ``um+2 s-1'' stands for micrometers squared per second
\item
  ``60 s'' stands for a minute
\item
  ``10+3 m'' stands for a kilometer
\end{itemize}

\subsubsection{The ``SI'' unit system}

The ``SI'' unit system \citep{SI} defines SI base units (§2.1), SI
derived units (§2.2), and SI prefixes (§3.1).

\begin{longtable}[c]{@{}lll@{}}
\hline\noalign{\medskip}
dimension
 & symbol
 & unit name
\\\noalign{\medskip}
\hline\noalign{\medskip}
length
 & m
 & meter
\\\noalign{\medskip}
mass
 & kg
 & kilogram
\\\noalign{\medskip}
time
 & s
 & second
\\\noalign{\medskip}
electric current
 & A
 & ampere
\\\noalign{\medskip}
temperature
 & K
 & kelvin
\\\noalign{\medskip}
amount of substance
 & mol
 & mole
\\\noalign{\medskip}
luminous intensity
 & cd
 & candela
\\\noalign{\medskip}
\hline
\noalign{\medskip}
\caption{SI base unit symbols and names.}
\end{longtable}

\begin{longtable}[c]{@{}llll@{}}
\hline\noalign{\medskip}
dimension
 & symbol
 & unit name
 & conversion
\\\noalign{\medskip}
\hline\noalign{\medskip}
plane angle
 & rad
 & radian
 & 1 rad = 1 m m⁻¹
\\\noalign{\medskip}
solid angle
 & sr
 & steradian
 & 1 sr = 1 m² m⁻²
\\\noalign{\medskip}
frequency
 & Hz
 & hertz
 & 1 Hz = 1 s⁻¹
\\\noalign{\medskip}
force
 & N
 & newton
 & 1 N = 1 m kg s⁻²
\\\noalign{\medskip}
pressure/stress
 & Pa
 & pascal
 & 1 Pa = 1 N m⁻²
\\\noalign{\medskip}
energy/work
 & J
 & joule
 & 1 J = 1 N m
\\\noalign{\medskip}
power
 & W
 & watt
 & 1 W = 1 J s⁻¹
\\\noalign{\medskip}
electric charge
 & C
 & coulomb
 & 1 C = 1 A s
\\\noalign{\medskip}
voltage
 & V
 & volt
 & 1 V = W A⁻¹
\\\noalign{\medskip}
capacitance
 & F
 & farad
 & 1 F = C V⁻¹
\\\noalign{\medskip}
electric resistance
 & ohm
 & ohm
 & 1 Ω = V A⁻¹
\\\noalign{\medskip}
electric conductance
 & S
 & siemens
 & 1 S = 1 A V⁻¹
\\\noalign{\medskip}
magnetic flux
 & Wb
 & weber
 & 1 Wb = 1 V s
\\\noalign{\medskip}
magnetic flux density
 & T
 & tesla
 & 1 T = 1 Wb m⁻²
\\\noalign{\medskip}
inductance
 & H
 & henry
 & 1 H = 1 Wb A⁻¹
\\\noalign{\medskip}
Celsius temperature
 & degC
 & degree Celsius
 & 0 °C = 273.15 K
\\\noalign{\medskip}
luminous flux
 & lm
 & lumen
 & 1 lm = 1 cd sr
\\\noalign{\medskip}
illuminance
 & lx
 & lux
 & 1 lx = 1 lm m⁻²
\\\noalign{\medskip}
radioactivity
 & Bq
 & becquerel
 & 1 Bq = 1 s⁻¹
\\\noalign{\medskip}
absorbed dose
 & Gy
 & gray
 & 1 Gy = 1 J kg⁻¹
\\\noalign{\medskip}
dose equivalent
 & Sv
 & sievert
 & 1 Sv = 1 J kg⁻¹
\\\noalign{\medskip}
catalytic activity
 & kat
 & katal
 & 1 kat = 1 mol s⁻¹
\\\noalign{\medskip}
\hline
\noalign{\medskip}
\caption{SI derived unit symbols, names and conversion rules.}
\end{longtable}

\begin{longtable}[c]{@{}lll@{}}
\hline\noalign{\medskip}
prefix
 & symbol
 & factor
\\\noalign{\medskip}
\hline\noalign{\medskip}
exa-
 & E
 & 10¹⁸
\\\noalign{\medskip}
peta-
 & P
 & 10¹⁵
\\\noalign{\medskip}
tera-
 & T
 & 10¹²
\\\noalign{\medskip}
giga-
 & G
 & 10⁹
\\\noalign{\medskip}
mega-
 & M
 & 10⁶
\\\noalign{\medskip}
kilo-
 & k
 & 10³
\\\noalign{\medskip}
hecto-
 & h
 & 10²
\\\noalign{\medskip}
deca-
 & da
 & 10¹
\\\noalign{\medskip}
deci-
 & d
 & 10⁻¹
\\\noalign{\medskip}
centi-
 & c
 & 10⁻²
\\\noalign{\medskip}
milli-
 & m
 & 10⁻³
\\\noalign{\medskip}
micro-
 & u
 & 10⁻⁶
\\\noalign{\medskip}
nano-
 & n
 & 10⁻⁹
\\\noalign{\medskip}
pico-
 & p
 & 10⁻¹²
\\\noalign{\medskip}
femto-
 & f
 & 10⁻¹⁵
\\\noalign{\medskip}
atto-
 & a
 & 10⁻¹⁸
\\\noalign{\medskip}
\hline
\noalign{\medskip}
\caption{SI prefixes.}
\end{longtable}

\twocolumn

\bibliographystyle{elsarticle-num}
\providecommand{\urlprefix}{} % suppress printing of "URL"
\bibliography{h5md}

\end{document}